# Plasmon-enhanced optical control of magnetism at the nanoscale via the inverse Faraday effect

Sergii Parchenko[1,2,3]*, Kevin Hofhuis[1,2,4], Agne Ciuciulkaite[5,$], Vassilios Kapaklis[5], Valerio Scagnoli[1,2], Laura Heyderman[1,2], Armin Kleibert[6]*

1. Laboratory for Mesoscopic Systems, Department of Materials, ETH Zurich, 8093 Zurich, Switzerland
2. Laboratory for Multiscale Materials Experiments, Paul Scherrer Institute, 5232 Villigen PSI, Switzerland
3. European XFEL, Holzkoppel 4, 22869 Schenefeld, Germany
4. Laboratory for Nano and Quantum Technologies, Paul Scherrer Institute, 5232 Villigen PSI, Switzerland
5. Department of Physics and Astronomy, Uppsala University, Box 516, 751 20 Uppsala, Sweden
6. Swiss Light Source, Paul Scherrer Institute, 5232 Villigen PSI, Switzerland

*Corresponding authors: sergii.parchenko@xfel.eu, armin.kleibert@psi.ch

**The relationship between magnetization and light has been the subject of intensive research for the past century, focusing on the impact of magnetic moments on light polarization. Conversely, the manipulation of magnetism through polarized light is being investigated to achieve all-optical control of magnetism in spintronics. While remarkable discoveries such as single pulse all-optical switching of the magnetization in thin films and sub-micrometer structures have been reported, the demonstration of local optical control of magnetism at the nanoscale has remained elusive. Here, we show that exciting gold nanodiscs with circularly polarized femtosecond laser pulses leads to the generation of sizeable local magnetic fields that enable ultrafast local control of the magnetization of an adjacent magnetic film. In addition, we find that the highest magnetic fields are generated when exciting the sample at a wavelength larger than that of the actual plasmonic resonance of the gold nanodiscs, so avoiding undesired heating effects due to absorption. Our study paves the way for light-driven control in nanoscale spintronic devices and provides important insights into the generation of magnetic fields in plasmonic nanostructures.**

The manipulation of magnetic states in materials using light has been the subject of intense research for several decades[1,2]. With the development of high-intensity pulsed light sources, it is now possible to modify the magnetization within time scales comparable to the fundamental electronic interaction times in magnetic materials such as electron-electron, electron-spin and electron-lattice interactions. Importantly, femtosecond laser pulses can be used to rapidly quench the magnetization within a sub-picosecond time frame due to ultrafast heating[3]. Here, the ultra-short laser pulse drives the material to an out-of-equilibrium state by decoupling the spin, electron, and lattice system during the excitation process, inducing ultrafast

---

demagnetization dynamics[4]. Another remarkable effect of the interaction of ultra-short pulses with magnetic materials is all-optical magnetization switching where the magnetization state can be reversed at the picosecond time scale by means of photoexcitation[5]. Currently, all-optical magnetization switching is considered to be primarily driven by the heating effects[6], and a variety of materials and engineered structures are known to undergo all-optical magnetization reversal either as a stochastic multi-pulse process or single pulse toggle switching[7,8,9].

Another way to change the magnetization state with light is through the inverse Faraday effect (IFE), where circularly polarized light acts as an effective magnetic field that changes the net magnetic moment of the system[10,11]. The sign of the effective magnetic field generated by light due to the IFE depends on the handedness of the circular polarization[12], hence, making it a promising approach for ultrafast opto-spintronic applications[13,14]. Indeed, in several studies, it has been shown that magnetization precession can be triggered by the IFE[12,13,15,16,17], and the role of the IFE in all-optical magnetization switching with single and multiple pulses is the focus of ongoing theoretical and experimental research[18,19,20,21,22].

Despite the impressive progress in understanding the interaction of ultrafast laser pulses with magnetic matter and the technological potential to achieve ultrafast, contactless control of magnetism, there are important challenges for the implementation of all-optical magnetization control in real applications. First of all, for most applications it will be necessary to address nanoscale structures that are much smaller than the optical wavelength of the exciting laser sources[23]. Indeed, heat-assisted switching was demonstrated on the micrometre[24] and nanometre scales[25,26], but deterministic control of the magnetization on the nanoscale is yet to be achieved. Another challenge is that a strong magneto-optical response is typically required to generate a substantial magnetic moment by the IFE, but achieving this is not easy because the associated strong optical absorption is typically unfavourable for the IFE. This is because the IFE field in metals is generated when the photon's electromagnetic field acts on the medium's electrons, inducing electron motion in the direction of the electric field associated with the incoming illumination[27,28]. If the photon is absorbed, however, this effect is suppressed. For this reason, in previous experimental demonstrations of large amplitude magnetization dynamics driven by the IFE, the focus has mainly been on inducing this effect in magnetic dielectrics with strong magneto-optical response butusing optical excitation below the bandgap. Despite the challenges associated with achieving the generation of a substantial magnetic moment with an optical pulse, IFE is an active area of research and development, with potential benefits for prospective devices, including lower energy requirements and increased operation speed.

One way to achieve effective control of the magnetization at the nanoscale is to employ plasmonic nanostructures. These structures can be used to concentrate the electromagnetic field of incident light down to the nanoscale[29,30], resulting in a higher efficiency of coupling between the light and the material near the nanostructures. This has opened up new possibilities for achieving effective control of magnetization at the nanoscale using light. Moreover, when using circularly polarized light and suitable geometries, plasmonic nanostructures have been predicted to generate effective magnetic fields with magnitudes of the order of one Tesla[23,31,32,34], and recent experimental demonstrations using plasmonically active gold nanoparticles have generated a magnetic moment of approximately 0.95 $\mu_B$ per gold atom[33]. Here, the magnetic moment originates from the motion of the electrons, which follow the electric field of the incident laser pulse. However, coupling between the light and the plasmonic mode leads to increased absorption and a temperature increase, which can limit the effectiveness of this approach for the nanoscale devices. To overcome this limitation, it is necessary to explore ways to generate a significant magnetic moment in nanostructures while avoiding increased absorption. Here, we demonstrate that it is possible to achieve a significant enhancement of the IFE in the vicinity of plasmon resonance at the nanoscale without increasing absorption. This paves the way towards developing novel, high-performance nanoscale devices based on the IFE.

The system under examination consists of plasmonic gold nanodisc arrays on top of a ferrimagnetic $Tb_{24}Co_{76}$ alloy film with out-of-plane magnetization[35] (see Fig. 2a-c and Methods section). First, we address the possibility of local IFE amplification in the vicinity of the localized surface plasmon resonance (LSPR) [36], in arrays of nanodiscs with radius $R$ varying from 30 nm to 120 nm and for different excitation wavelengths ranging from $\lambda = 300$ nm to 1300 nm using finite-difference time-domain (FDTD) calculations. The simulations include circularly polarized light at perpendicular incidence, as schematically shown in the inset of Fig. 1a, and periodic boundary conditions with 50 nm edge-to-edge spacing between the nanodiscs to mimic the experimentally investigated samples (see Methods section for more details). The efficiency of opto-plasmonic coupling depends on both the radius of the nanodiscs, $R$, and the wavelength of the incoming light, $\lambda$. In Fig. 1a, we show the excitation efficiency of LSPR in nanodiscs[36] as a function of $\lambda$ and $R$. Here we see that the $\lambda$ where the LSPR is excited, which corresponds to the maximum absorption shown by the dashed red line in Fig. 1a, linearly increases with increasing $R$.

Shown in panels b – e of Fig. 1 is the calculated spatial distribution of electric field $E$ and magnetic field $H$ generated by circularly polarized light with $\lambda = 1030$ nm in nanodiscs for $R = 50$ and 100 nm. For $R = 50$ nm, when the optical excitation matches the LSPR mode, there is a strong concentration of the electric field $E$ at the edges of gold nanodiscs (Fig. 1b) and some concentration of magnetic field $H$ at the interface between the plasmonic nanodiscs and the magnetic thin film (Fig. 1d). The observed distribution of the $E$ and $H$ field is as expected for plasmon excitation and suggests an increased absorption. In contrast, for larger nanodiscs with $R = 100$ nm, there is a non-resonant excitation and only a small concentration of the $E$ field but with a strong enhancement of the $H$ field. This means that there is no significant absorption increase when nanodiscs with $R = 100$ nm are optically excited, but the plasma electrons can still be driven by circularly polarized light pulses and a magnetic moment can be generated that is potentially larger than the magnetic moment generated with on-resonant plasmonic excitation.

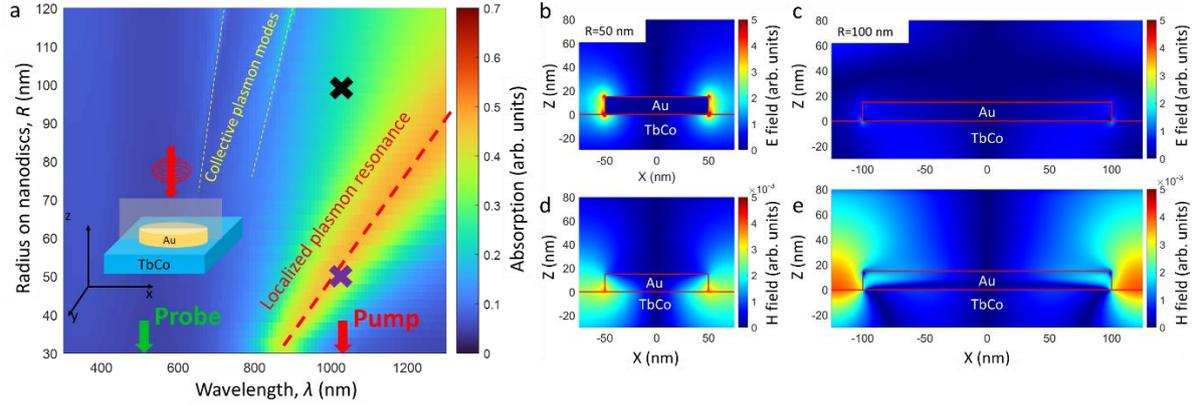

*Figure 1. FDTD calculations of LSPR excitation with circularly polarized light.* (a) 2D map of the absorption on changing the excitation laser wavelength, $\lambda$, and radius of the nanodiscs, R. The inset on the left side shows the geometry used for the calculations. The red dashed line indicates the localized surface plasmon resonance branch. The yellow dashed lines indicate collective plasmonic modes. Purple and black crosses indicate the $\lambda$ and R used in panels (b), (d) and (c), (e), respectively. The wavelength of the pump and probe laser beams used in the time-resolved experiments are indicated with red and green arrows, respectively. (b)-(e) Spatial distribution of electric field E (b), (c) and magnetic field H (d), (e) for nanodiscs with R = 50 nm (b), (d) and with R = 100 nm (c), (e) generated in plasmonic nanodiscs when excited with circularly polarized light with $\lambda$=1030 nm.

In order to confirm the optical generation of magnetic moments in nanostructures away from plasmon resonance we perform experimental time-resolved studies using a two-colour pump-probe approach. Traditional LSPR investigations typically involve selecting a specific plasmonic nanoantenna design while probing the optical and magnetic response of the sample as a function of wavelength. The drawback to this method is that the optical response of the material also varies with the wavelength, which may mask plasmon-induced effects. Instead, we perform experiments at a single excitation wavelength and control the efficiency of the LSPR excitation by altering the size of the plasmonic nanostructure. In this way we tune the efficiency of LSPR excitation at a single chosen wavelength, thus isolating the effects of plasmon resonance excitation from the material response. Scanning electron microscope micrographs of samples with nanodisc radii $R = 30$ nm and $R = 100$ nm are shown in Fig. 2a,b, respectively. The constant 50 nm edge spacing between the individual nanodiscs ensures highly dense arrays of nanodiscs covering 45 – 55% of the TbCo film surface depending on the disc radius. Thus, a large proportion of the TbCo film will be affected when exciting the nanodiscs, increasing the detectable magnetic signal. At the same time, the chosen experimental parameters avoid excitation of collective plasmonic modes at the pump wavelength (see Fig. 1a), which could otherwise couple to the LSPR mode[37].

The experimental geometry is depicted in Fig. 2c and consists of ultrashort laser pulses with circular polarization and $\lambda = 1030$ nm for optical excitation and probe pulses with linear polarization and $\lambda = 515$ nm. It is known that the excitation of the plasmon resonance may affect the polarization state of the light[38,39]. Bearing this in mind, if we measure the magnetization dynamics by analysing the change in the polarization state of the probe light, the change in the light polarization due to the plasmon excitation may mask plasmon-induced changes to the magnetization state. To minimize this effect, the size of the nanodiscs was chosen to ensure that the probe beam does not excite any plasmon modes, see Fig. 1a, thus allowing us to only probe the dynamics in the TbCo film. Both the excitation and probe beams are incident on the sample at a normal angle and have a pulse duration of approximately 250 fs. The magnetization change is tracked by measuring the time-resolved rotation of the polarization plane of the transmitted probe beam due to the direct Faraday effect using a balanced detection scheme. The pump fluence during the experiments was kept at $\Phi = 0.5$ mJ/cm$^2$ to prevent damage to the nanostructures, which would occur at higher fluences. The experiments were conducted in the absence of an external magnetic field. More information on the experimental procedure are available in the Methods section.

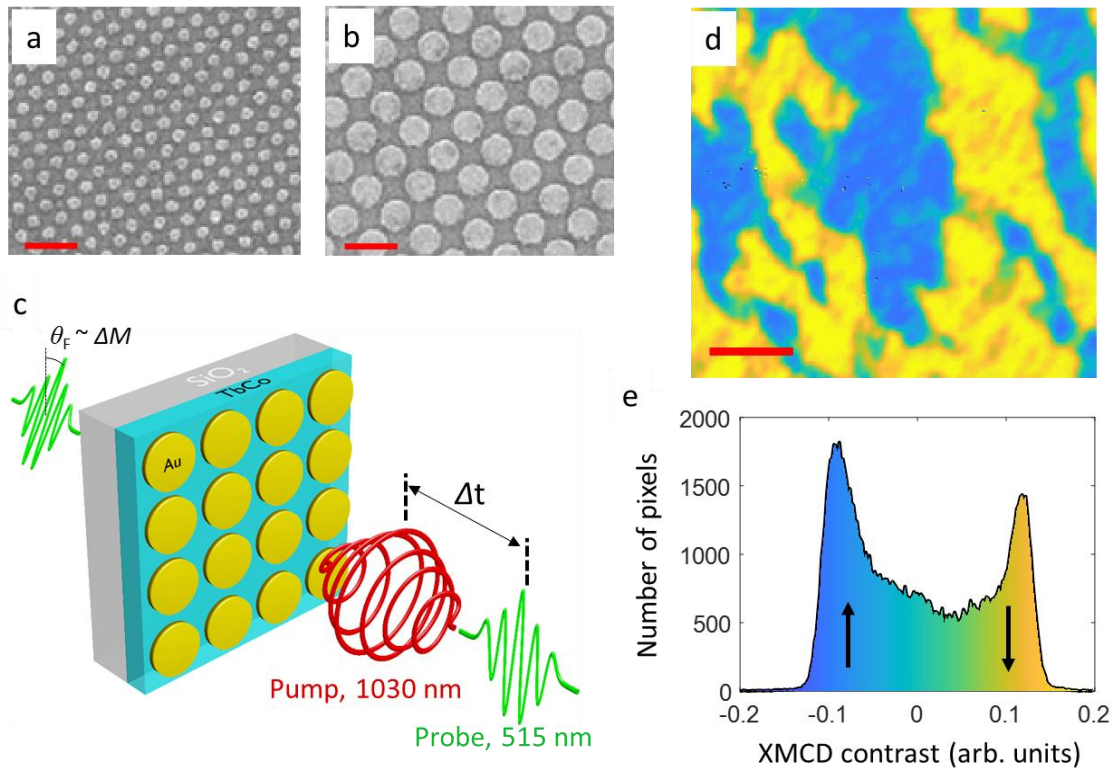

*Figure 2. Sample structure and geometry of the time-resolved experiments on TbCo with patterned gold nanodiscs.* *(a) and (b) Scanning electron microscopy images of Au nanodiscs on top of a TbCo film with R=30 nm and R=100 nm, respectively. The red scale bar on both images corresponds to 300 nm. The spacing between the individual nanodiscs is 50 nm for both arrays. (c) Experimental geometry for the time-resolved magnetization dynamics experiments. The sample was excited with a circularly polarized λ = 1030 nm pump beam and the magnetization change was tracked following the Faraday rotation of a linearly polarized λ = 515 nm probe beam. (d) A representative magnetic domain pattern in the TbCo film visualized taking advantage of the x-ray magnetic circular dichroism (XMCD) effect at the Co L$_3$ edge recorded by means of x-ray photoemission electron microscopy. Blue and yellow domains correspond to regions with the magnetization pointing upwards or downwards, respectively. The red scale bar corresponds to 2 μm (e) Histogram of the XMCD contrast values derived from the image in panel (d).*

In our experiments, we aim to exploit the plasmon resonance to modify the magnetic state in the thin-film/nanodisc system. However, the magnetic circular dichroism effect associated with the TbCo film might bring an additional contribution to the measured signal, masking the plasmon-induced dynamics[40]. The different absorption for left- and right-handed circularly polarized pump means that, for a uniformly magnetized sample, the demagnetization rate for the different polarizations will not be the same. This would result in helicity-dependent light-induced modification of the magnetic state and will contribute to the signal when comparing the dynamics excited with different polarization of pump pulses. To avoid this issue, we make sure that the TbCo film used in the experiment is in a multidomain state. Indeed, magnetic contrast maps of the TbCo film, recorded with XMCD PEEM, reveal irregular domains with magnetization pointing up (in blue) and down (in yellow), as shown in Fig. 2d (see Methods for more information about the experimental procedure). The typical domain size is of the order of several micrometers, so much smaller than the pump and probe laser spot footprint of 200 µm, so avoiding the situation where large magnetic domains would result in dichroism-related effects.

A detailed statistical analysis of the image pixels shows an imbalance in the extent of up- and down magnetized areas (Fig. 1e). The magnetic domains with up (down) orientation of magnetization give positive (negative) contributions to the Faraday rotation of the probe beam and, for a fully demagnetized sample, the contributions of both types of domains would be equal and it would not be possible to measure the demagnetization dynamics expected on femtosecond pulse excitation. However, due to the small imbalance of "up domains" and "down domains", we can still detect the pump-induced demagnetization dynamics, but at the same time, the measured magnetic signal is almost independent of the magnetic circular dichroism effect, which has an opposite contribution for areas with magnetization pointing up and down. It should also be noted that while the gold nanodiscs do not appear in the magnetic contrast images, we have observed no spatial correlation between the magnetic domains and the gold nanodisc arrays.

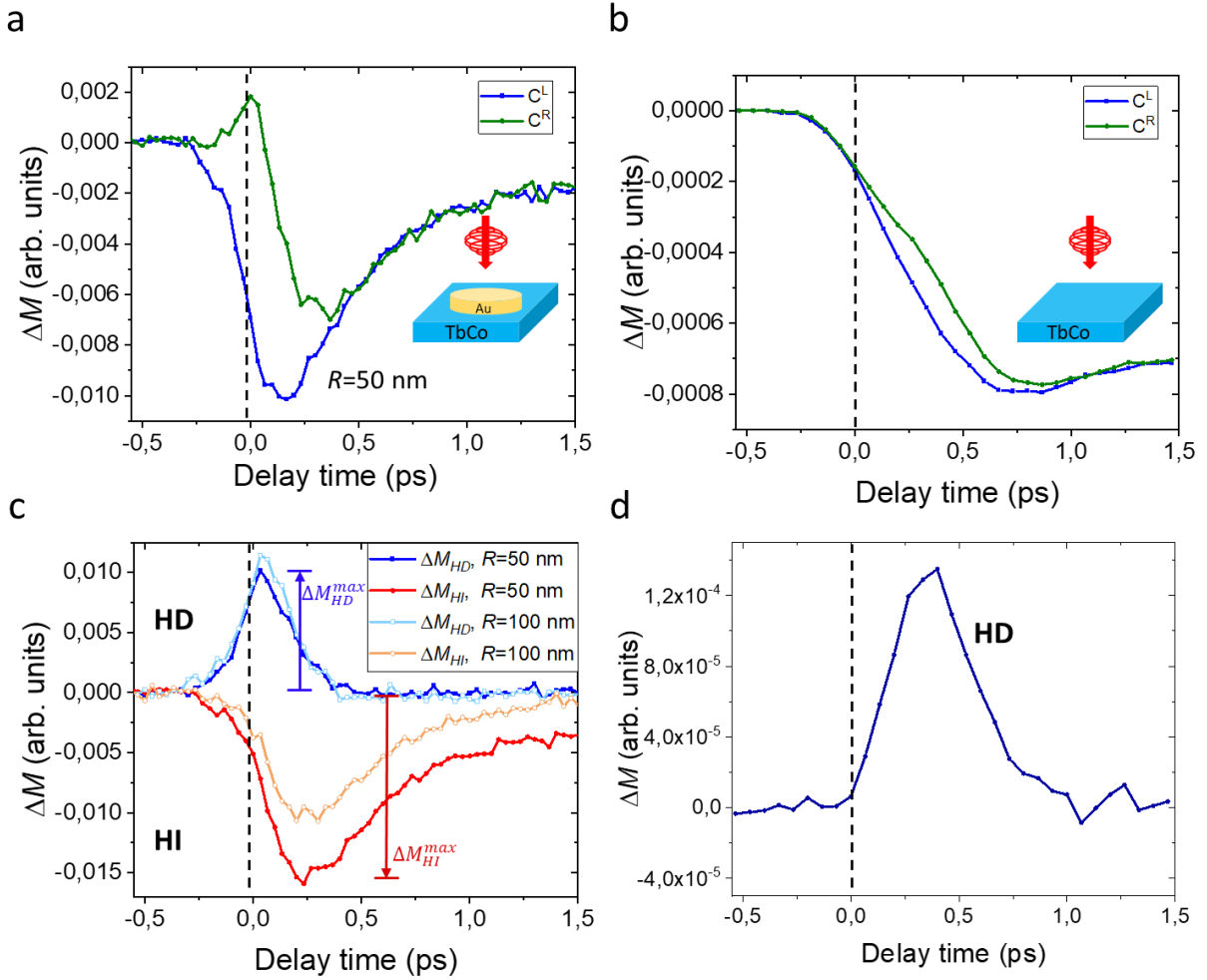

*Figure 3. Magnetization dynamics following laser excitation of TbCo in the presence of gold plasmonic structures.* Time-resolved magnetization changes after the pump excites the part of the sample with arrays of gold nanodiscs with R = 50 nm (a) and the bare TbCo film (b). The difference between traces measured after excitation with opposite circular polarization of pump light is due to the IFE. (c) Helicity-dependent (HD) and helicity-independent (HI) components of the magnetization change, $\Delta M$, for nanodiscs with R = 50 nm and R = 100 nm. HD and HI signals have different dependencies for nanodiscs with different radii. The maximum values of each component discussed in the text are indicated. (d) HD component for bare TbCo film. Note that the maximum intensity of laser-induced magnetic field occurs at a later time delay compared to nanostructures.

The pump-induced change of magnetization, $\Delta M$, for the two opposite chiralities of the circularly polarized pump light, tracked with the probe beam, is shown in Fig. 3. Besides the thermally induced demagnetization in the TbCo film, we observe a distinct evolution of $\Delta M$ when pumped with opposite circular polarization for the sample with nanostructures (Fig. 3a) and the bare TbCo film (Fig. 3b). The distinct magnetization dynamics at early delay times, when excited with opposite chirality of circular polarization, is a footprint of the IFE, as the sign of magnetic moment generated by photons depends on the sign of the circular polarization. In case of the bare TbCo film, these observation confirms recent theoretical predictions of the intrinsic IFE at the pump wavelength[35]. In case of the sample with the gold nanodiscs we assign these data to a combination of the intrinsic IFE and the plasmonic-assisted IFE acting on the TbCo as discussed further below.

In order to discriminate non-thermal and thermal effects, we extract two components from the experimental time traces: the helicity-dependent component $\Delta M_{HD}$ that is associated with the optical generation of the magnetic moment in the thin-film/nanodisc system and the helicity-independent component $\Delta M_{HI}$ that is associated with the pump-induced demagnetization process in TbCo [40]. These components can be represented as follows:

$$\Delta M_{HD}(t) = \Delta M_{C^L}(t) - \Delta M_{C^R}(t)$$

$$\Delta M_{HI}(t) = \Delta M_{C^L}(t) + \Delta M_{C^R}(t)$$

where $\Delta M_{C^L}(t)$ and $\Delta M_{C^R}(t)$ are the time-resolved magnetization dynamics when pumped with left- and right-handed circular polarization, respectively. The HD and HI components of the dynamics in the TbCo film with nanodiscs and the HD component of the dynamics in the bare TbCo film are plotted in Fig. 3c,d, respectively. The HD component of the dynamics in the TbCo film with nanodiscs arises immediately when the pump light reaches the material (see blue curves in Fig. 3c), while the maximum light-generated magnetic moment for the bare TbCo film without the nanodiscs arises at about 0.4 ps after the excitation. An explanation for this difference is that the light-generated magnetic moment for the nanostructures originates from the circular drift motion of plasma electrons in the Au nanodiscs that follow the electric field component of the pump light and starts immediately when the light reaches the sample. When the bare film is excited, the angular momentum from the pump pulse is first absorbed by electrons and only later transferred to the spins, a process that is mediated by the electron-spin interaction. A similar delay in the HD magnetization evolution in TbCo has been reported previously[40]. Furthermore, comparing the ratio of respective maximum values $\Delta M_{HD}^{max}/\Delta M_{HI}^{max}$ for the TbCo film with and without nanodiscs, we see that optically generated magnetization is about an order of magnitude stronger at the TbCo film with the nanostructures compared to the bare film, demonstrating a significant increase in the IFE-generated magnetic moment in the presence of nanodiscs.

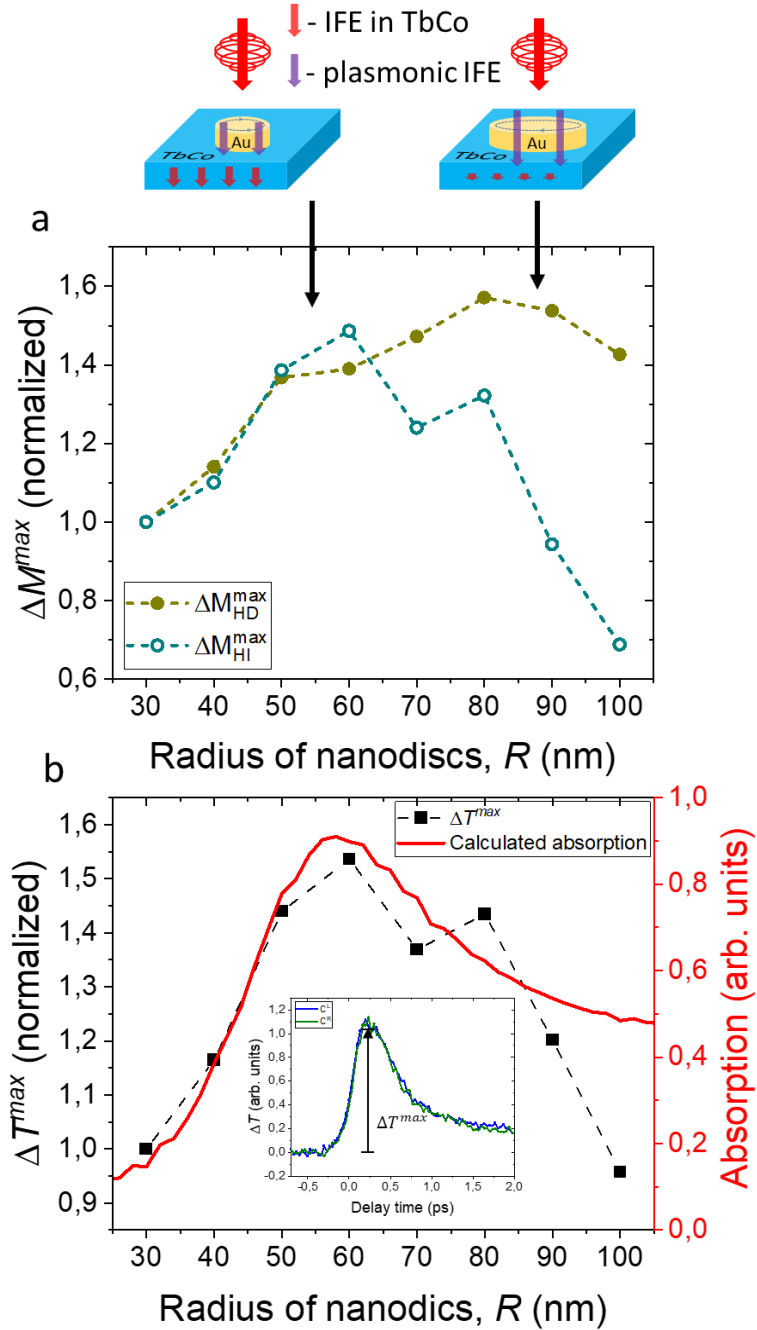

*Figure 4. **Comparison of different components of the magnetic signal as a function of nanodisc radius** (a) Maximum of $\Delta M_{HD}$ and $\Delta M_{HI}$ components of the laser-induced magnetization change a function of the radius of nanodiscs. The data are normalized to their value for R=30 nm for better presentability. The schematics above panel (a) indicate the different contributions to the $\Delta M_{HD}$ signal for nanostructures with different radii. Purple and red arrows represent contributions from plasmonic IFE and IFE in the TbCo film, respectively. (b) Comparison of maximum pump-induced transmittance changes $\Delta T^{max}$ as a function of the radius of nanodiscs, R, and calculated absorption coefficient. The data for $\Delta T^{max}$ is normalized to its value for R=30 nm. The inset to panel b shows an example of a transient transmittance signal after the pump excites the part of the sample with arrays of gold nanodiscs with R = 50 nm.*

It is generally assumed that the plasmon-assisted action of light on the material is driven by the concentration of the electric field of incoming light at the nanostructures, seen as an increased absorption when conditions for plasmon resonance excitation are met[41]. In our case, there are three contributions to the plasmon-assisted optical modification of the magnetic state in the thin-film/nanodisc system after the optical excitation. First, the pump-induced IFE in the TbCo film generates a magnetic moment in the sample by virtue of the enhanced absorption of the incoming light due to coupling to the plasmon resonance, resulting in a stronger electric field inside the material (see Fig. 1b). Second, the light-induced drift current of the electrons in the plasmonic nanodiscs creates an effective magnetic moment – plasmonic IFE that acts on the magnetic state of the TbCo film. The plasmonic IFE is also expected to be enhanced near the plasmon resonance, which allows for a more efficient localization of the electric field from the pump beam in the plasmonic nanostructures and this should reach maximum at the plasmon resonance. The third mechanism is an increased optically-induced demagnetization due to plasmon-mediated increased absorption when the plasmon resonance is excited. Thus, it is expected that the effective magnetic moment generation through the circular motion of electrons in the nanodiscs and the IFE enhancement in the TbCo film (both of which contribute to the $\Delta M_{HD}$), and the light-induced thermal demagnetization process ($\Delta M_{HI}$ component) are all proportional to the increased absorption due to the plasmon resonance excitation. However, the data in Fig 3c suggests otherwise. Surprisingly, the $\Delta M_{HD}^{max}$ for the TbCo film with $R = 100$ nm Au nanodisc arrays is slightly larger than $\Delta M_{HD}^{max}$ for the case of $R = 50$ nm, while the $\Delta M_{HI}^{max}$ for $R = 100$ nm is about 30% lower than $\Delta M_{HI}^{max}$ for $R = 50$ nm.

To understand this discrepancy, we compare the $\Delta M_{HD}^{max}$ and $\Delta M_{HI}^{max}$ values as a function of nanodisc radius $R$ that are shown in Fig 4a. For the sake of clarity, the curves are normalized to their maximum value at $R = 30$ nm. We find that for radii up to $R = 60$ nm, the amplitude of both $\Delta M_{HD}$ and $\Delta M_{HI}$ components increases in a similar way. However, for $R > 60$ nm, the $\Delta M_{HI}^{max}$ decreases while $\Delta M_{HD}^{max}$ continues to increase. Furthermore, in the graph in Fig. 4b the amplitude of transient transmittance change $\Delta T^{max}$ at our probe wavelength (where no plasmonic resonance is excited - see dispersion of plasmon modes in Fig. 1a) is given for different $R$. In particular, the $\Delta T^{max}$ dependence in Fig. 4b reflects the pump induced change in the electronic system when pumping the TbCo film with nanodiscs with different radii. The dependence of $\Delta M_{HI}^{max}$ as a function of $R$ is found to be very similar to $\Delta T^{max}$ (compare data in Fig. 4a and b), which can be understood from the fact that both $\Delta M_{HI}$ and $\Delta T^{max}$ dependencies as a function of $R$ reflect the increasing absorption due to coupling between the pump beam and the localized plasmon resonance in the gold nanodiscs. We found that the optimal conditions for the plasmonic resonance excitation with a pump beam $\lambda = 1030$ nm were realized for nanodiscs with radii of $R = 50 - 60$ nm. Here, the $\Delta M_{HI}^{max}$ and $\Delta T^{max}$ are maximal. Under these conditions, a magnetic moment is generated in the thin-film/nanodisc system due to both plasmonic IFE in the Au nanodiscs that acts on the TbCo film and the IFE in the TbCo film with increased absorption. The calculation of optical absorption by nanodiscs of different radii for $\lambda = 1030$ nm, as shown by the red curve in Fig. 4b, also shows that the maximal absorption is at $R = 55$ nm. Strikingly, the $R$-dependence of the $\Delta M_{HD}^{max}$ value (Fig. 4a) suggests that plasmonic IFE enhancement does not rely on absorption because the $\Delta M_{HD}^{max}$ is largest at a radius of $R = 90 – 100$ nm where the plasmonic excitation is off resonant. In this case, most of the magnetic moment is generated in the plasmonic nanostructures at reduced absorption. Nevertheless, there is still a fraction of the IFE-generated magnetization in the TbCo film, with an efficiency that is comparable with the effect seen in the bare film (see Fig. 3b). The amplitude of the different contributions to the $\Delta M_{HD}$ signal when pumping nanodiscs with different radii are depicted in the schematics above Fig. 4a with purple and red arrows.

In summary, we have demonstrated that plasmonic nanostructures can be exploited to increase the opto-magnetic response and generate a large effective magnetic field through the plasmonic IFE. This opens up new possibilities for the manipulation and control of magnetic materials using light, which has numerous implications for the development of next-generation data storage, data processing and sensing devices. The observation that plasmonic nanostructures can generate a strong effective magnetic field, even when the conditions for plasmon resonance excitation are not perfectly met, is particularly interesting. Specifically, this provides a means to efficiently generate a magnetic field while avoiding a significant increase in absorption, which is a major advantage for heat-sensitive applications. For example, there are several approaches proposed to manipulate the state of quantum systems that rely on the effective magnetic field generated by light for controlling superconducting quantum bits[42,43] and in controlling photonic networks that operate on neuromorphic principles[44]. Here, a minimization of absorption to avoid decoherence would be highly beneficial. The striking observation that plasmonic IFE is much stronger slightly away from the plasmon resonance implies that further unexpected plasmon-related effects might be discovered when conditions for plasmon resonance excitation are not perfectly met. Such unexpected phenomena would lead the way to novel applications involving the opical control of quantum states on the nanoscale.

**Acknowledgements:**

We gratefully thank Anja Weber from PSI for her support during the preparation of nanostructures. KH and LH acknowledge funding from the Swiss National Science Foundation (Project No. 200020_172774). This work is part of a project which has received funding from the European Union's Horizon 2020 research and innovation program under grant agreement No. 737093, "FEMTOTERABYTE". We gratefully acknowledge the support of the COST Action CA17123 MAGNETOFON.

## Methods.

**Sample preparation**. The sample consists of a set of several 100 μm × 100 μm arrays of gold nanodiscs with different radii on top of a TbCo film. An amorphous 20 nm - thick film with a nominal composition of $Tb_{24}Co_{76}$ was sputter deposited with DC magnetron using Tb and Co targets in an argon atmosphere onto a $SiO_2$ substrate. A 2 nm - thick $Al_2O_3$ capping layer was deposited on top of the TbCo film to prevent oxidation. The thin film preparation method, as well as the structural and magnetic characterizations, are described in detail in Ref. [35]. TbCo is a ferrimagnetic alloy with two antiparallel magnetic sublattices made of Tb and Co. For the given concentration of magnetic elements, the film has an out-of-plane magnetic anisotropy with a coercive field around $H_C$ = 800 mT. The relatively high $H_C$ is favourable for our experiments as the magnetic domain structure is more robust to optical excitation. Arrays of gold nanodiscs with radii varying from $R$ = 30 nm to $R$ = 100 nm and with a thickness of 15 nm were prepared using electron-beam lithography and thermal evaporation with subsequent lift-off.

**FDTD calculations**. Numerical calculations of the optical response of plasmonic nanodiscs were performed with Ansys Lumerical FDTD software. The structure used in the calculation reflected the actual sample geometry and consisted of TbCo (20 nm)/$Al_2O_3$ (2 nm)/ Au nanodisc (15 nm). A schematic of the system geometry used for the calculations is shown in the inset of Fig. 1a. The source of circularly polarized light was obtained from a combination of two sources of linearly polarized light with orthogonal polarization direction and a phase shift of π/2. We used Bloch boundary conditions in the x and y direction, which are perpendicular to the propagation of circularly polarized light. Along the z direction, which is parallel to the propagation direction of light, we used perfectly matching layer (PML) boundary conditions. The cell size was chosen to be 2×2×2 $nm^3$.

**XPEEM characterization.** Magnetic domain imaging was carried out at the Surface/Interface: Microscopy beamline at the Swiss Light Source using the PEEM end station. Magnetic contrast was extracted using the XMCD effect by taking the asymmetry between two photoemission electron microscopy images recorded with left and right circularly polarized x-rays with x-ray photon energy $E$=779 eV, that correspond to Co $L_3$ absorption edge.

**Time-resolved experiments.** Optical pump-probe experiments to determine the magnetization dynamics were performed in transmittance (Faraday) geometry. Ultrashort laser pulses were generated with an ActiveFiber Yb pulsed fiber laser with a 200 kHz repetition rate and a fundamental wavelength of $\lambda$ = 1030 nm. The wavelength of the optical pump was $\lambda$ = 1030 nm with circularly polarized pulses. Linearly polarized probe pulses with $\lambda$ =515 nm were obtained by frequency doubling using a beta Barium Borate nonlinear crystal. The pump beam was focused to a spot size of 200 μm and the probe beam size was about 150 μm. Both beams propagate collinearly and impinge on the sample at the normal incidence. The pump fluence was 0.5 mJ/$cm^2$ and the probe beam fluence was about 1/50 of the pump fluence. Transient Faraday rotation of the probe beam was measured with a balanced detection scheme using a Zurich Instruments digital lock-in amplifier. Our study is based on comparing the time evolution of the magnetic signal obtained when pumping nanodiscs of different radii but with the same separation. This means that the area of the TbCo film covered with plasmonic nanodiscs of different radii varies, and the fraction of the probed area where the plasmonic nanodiscs alter the efficiency of optical generation of magnetic moment varies with the radius of the nanodiscs. In order to compensate for this, we correct the time traces to the fraction of the sample surface actually covered by nanostructures.

**Author contributions:** SP and AK conceived the research. AC and VK deposited the TbCo film, KH prepared the gold nanostructures, SP performed time-resolved experiments with support from VS, and SP performed numerical calculations. SP and AK performed the x-ray characterization. SP wrote the manuscript with the help of AK and LH and contributions from all co-authors.